\documentclass[preprint]{aastex}

\usepackage{float}

\usepackage{gensymb}

\usepackage{MnSymbol}

\usepackage[usenames,dvipsnames]{color}

\usepackage{natbib}

\shorttitle{PDS 144: Wide Binary}

\begin{document}

\title{PDS 144: The First Confirmed Herbig Ae - Herbig Ae Wide Binary}

\author{J.~B.~Hornbeck\altaffilmark{1}, C.~A.~Grady\altaffilmark{2,3}, M.~D.~Perrin\altaffilmark{4}, J.~P.~Wisniewski\altaffilmark{5,6}, B.~M.~Tofflemire\altaffilmark{5}, A.~Brown\altaffilmark{7}, J.~A.~Holtzman\altaffilmark{8}, K.~Arraki\altaffilmark{5,8}, K.~Hamaguchi\altaffilmark{9,3},  B.~Woodgate\altaffilmark{10}, R.~Petre\altaffilmark{10}, B.~Daly\altaffilmark{11,12}, N.~A.~Grogin\altaffilmark{4}, D.~G.~Bonfield\altaffilmark{10,13}, G.~M.~Williger\altaffilmark{1,14}, J.~T.~Lauroesch\altaffilmark{1}}

\altaffiltext{1}{Department of Physics \& Astronomy, University of Louisville, Louisville KY 40292 USA}
\altaffiltext{2}{Eureka Scientific, 2452 Delmer St. Suite 100, Oakland CA 96402, USA}
\altaffiltext{3}{Exoplanets and Stellar Astrophysics Laboratory, Code 667, Goddard Space Flight Center, 
Greenbelt MD 20771}
\altaffiltext{4}{Space Telescope Science Institute, Baltimore, MD 21218, USA}
\altaffiltext{5}{Department of Astronomy, University of Washington, Box 351580, Seattle, WA 98195}
\altaffiltext{6}{NSF Astronomy and Astrophysics Postdoctoral Fellow}
\altaffiltext{7}{CASA, University of Colorado, Boulder, CO 80309-0593, USA}
\altaffiltext{8}{Department of Astronomy, New Mexico State University, Dept 4500, Box 30001, Las Cruces, NM 88003}
\altaffiltext{9}{Department of Physics, UMBC, Baltimore, MD 21250, USA}
\altaffiltext{10}{NASA's Goddard Space Flight Center, Greenbelt MD 20771, USA}
\altaffiltext{11}{Department of Astronomy, University of Maryland, College Park, MD 20742-2421}
\altaffiltext{12}{NASA's GSFC Student Internship Program}
\altaffiltext{13}{Centre for Astrophysics Research, Science \& Technology Research Institute, University of Hertfordshire, College Lane, Hatfield, AL10 9AB, UK}
\altaffiltext{14}{Laboratoire Fizeau, Universit\'e de Nice, UMR 6525, 06108 Nice Cedex 2, France}

\begin{abstract}

PDS 144 is a pair of Herbig Ae stars that are separated by 5.35$\arcsec$ on the sky.  It has previously been shown to have an A2Ve Herbig Ae star viewed at 83$^\circ$ inclination as its northern member and an A5Ve Herbig Ae star as its southern member. Direct imagery revealed a disk occulting PDS 144 N - the first edge-on disk observed around a Herbig Ae star. The lack of an obvious disk in direct imagery suggested PDS 144 S might be viewed face-on or not physically associated with PDS 144 N.  Multi-epoch HST imagery of PDS 144 with a 5 yr baseline demonstrates PDS 144 N \& S are comoving and have a common proper motion with TYC 6782-878-1.  TYC 6782-878-1 has previously been identified as a member of Upper Sco sub-association A at d = 145 $\pm$ 2 pc with an age of 5 - 10 Myr.  Ground-based imagery reveals jets and a string of HH knots extending 13$\arcmin$ (possibly further) which are aligned to within 7$\degree$ $\pm$ 6$\degree$ on the sky. By combining proper motion data and the absence
of a dark mid-plane with radial velocity data, we measure the inclination of PDS 144 S to be i = 73$\degree$ $\pm$ 7$\degree$. The radial velocity of the jets from PDS 144 N \& S indicates they, and therefore their disks, are misaligned by 25$\degree$ $\pm$ 9$\degree$.  This degree of misalignment is similar to that seen in T-Tauri wide binaries.
\end{abstract}

Keywords: binaries: general - Protoplanetary disks - Stars: variables: T-Tauri, Herbig Ae/Be -  Stars: individual (PDS 144 N, PDS 144 S, TYC 6782-878-1) - Stars: pre-main sequence - ISM: Herbig-Haro objects  

\section{INTRODUCTION}

Most stars form in multiple star systems, the majority of which are binaries \citep{Mathieu_00}.  To date, studies of pre-main sequence binary systems have been largely composed of T-Tauri type stars because the Initial Mass Function limits the number of young massive stars like Herbig Ae/Be stars available for study \citep{Leinert_93, Ghez_93, White_01}.  Herbig Ae/Be stars are A or B stars with H$\alpha$ emission and identified by reflection nebulosity or other indicators of youth \citep{Herbig_60}.  Herbig Ae stars, like T-Tauri stars, are known to have circumstellar disks and have been observed driving jets \citep{Grady_00, Grady_04, Grady_10}. Relying on the assumption that Herbig Ae stars are the intermediate mass analogs to the T-Tauri stars, Herbig Ae stars in wide binary systems should display similar star-disk interactions as those seen in T-Tauri wide binaries.  We consider binary separations $>$ 100 AU as the definition of wide binary, as in \citet{Jensen_04}. The first possibility of a wide binary containing two Herbig Ae stars is PDS 144. \citet{Carballo_92} identified IRAS 15462-2551 with a previously anonymous 12th magnitude star in Scorpius. The Pico dos Dias Survey (PDS), a spectroscopic follow-up to Infrared Astronomical Satellite (IRAS ) sources \citep{Torres_95}, subsequently named this source PDS 144, and \citet{Perrin_06} found the two stars to be separated by 5.40\arcsec\ $\pm$ 0.01\arcsec. The northern member, PDS 144 N, is an A2Ve star with a disk viewed at 83$^\circ$ inclination from pole-on \citep{Perrin_06}.  The southern member, PDS 144 S, is an A5Ve star.  PDS 144 S also has an IR excess like PDS 144 N, but in direct imagery the diffracted and scattered light of PDS 144 S swamps any signal from the disk, and in the study by \citet{Perrin_06} it was unclear whether these two stars were in fact associated with one another.  \it Do these two stars make up the first confirmed Herbig Ae wide binary system? \rm 

\citet{Perrin_06} could not conclusively establish the physical association of these two stars, though their being a chance superposition is an unlikely scenario.  The only data previously available suggested a distance of 1 kpc for PDS 144 S and 2 kpc for PDS 144 N \citep{Vieira_03}.  Recent data from the Third U.S. Naval Observatory CCD Astrograph Catalog (UCAC3) suggest PDS 144 N \& S are not associated with one another because their data indicate oppositely directed proper motions \citep{Zacharias_10}.  However, the UCAC3 proper motions for PDS 144 N \& S are much larger (proper motion magnitude = 48 $\pm$ 27 mas yr$^{-1}$ and 112 $\pm$ 28 mas yr $^{-1}$ for PDS 144 N \& S respectively) than would be expected for stars at d $\geq$ 1 kpc.  \citet{Vieira_03} used only photometric data when estimating distances, and the low resolution of their observations prompted them to use the integrated magnitudes for the stars in PDS 144 due to their close proximity on the sky.  At the time it was unknown that PDS 144 N was completely occulted by its disk, drastically reducing its apparent luminosity in all wavelengths, and thus increasing its photometric distance.  A distance $\geq$ 1 kpc poses a problem in that it places both PDS 144 N \& S above the Galactic plane in an area devoid of any known star forming region or molecular cloud.   

In \S\ 2 of this paper we describe our observations of PDS 144 with the Hubble Space Telescope (HST) and ground based telescopes.  In \S\ 3  we will address the issue of association between PDS 144 N \& S, as well as determine age and distance for the system.  In \S\ 4 we place preliminary constraints on the inclination of the disk around PDS 144 S using data from both HST and ground-based instruments, and we summarize our results in \S\ 5.

\section{OBSERVATIONS \& DATA REDUCTION}

\subsection{Prime Focus Camera}

We obtained observations of PDS 144 on 2006 August 29 using the Prime Focus Camera (PFCam) on the Shane 3 m telescope at Lick Observatory.  Because PDS 144 was already setting at dusk, it was observed at an airmass of $\approx$3 and we were able to obtain just 750 s of exposure in a H$\alpha$ filter (6563/30 \AA ) and 500 s in [SII] (6720/30 \AA ), plus 90 s in a Spinrad R (R$_s$) filter for continuum (6580/1500 \AA ).  PFCam's detector is a Fairchild 4096$^2$ pixel CCD, its readout is binned 2 x 2 to yield 0.37$\arcsec$ pixel$^{-1}$ and it has a field of view (FOV) of 12.6$\arcmin$ on a side. After bias subtraction using the overscan regions, the images were registered together using cross-correlation of field stars, including compensation for field-dependent distortion introduced by the different thicknesses of the filters in the converging beam.  The distortion correction and astrometric solution were verified by comparison with 2MASS catalog stars in the field.   The R$_s$ image was then subtracted from each emission line image, with multiplicative scaling coefficients chosen to minimize the residual flux in a 512-pixel square subregion centered on the target. 

\begin{figure}[H]
\includegraphics[scale=1]{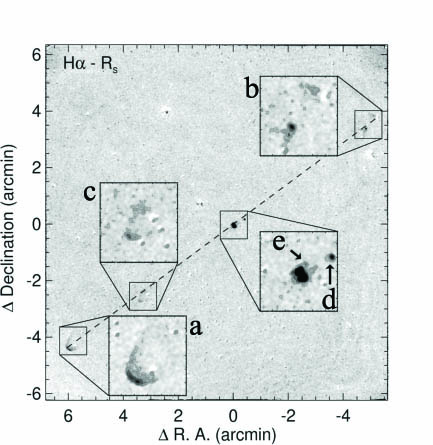}
\caption{\footnotesize PFCam image in H$\alpha$ centered on PDS 144 after subtracting the R$_s$ band image illustrating the multiple knots of HH 689 (a,b,c,d,e lettered radially inward toward the stars) associated with PDS 144 N \& S.   Lacking radial velocity information for the more distant HH knots, it is unclear to which star, PDS 144 N or S, HH 689a, b, \& c are associated. \normalsize \label{fig:Lickpds144_2}}
\end{figure}

In the resulting H$\alpha$ emission line image (Fig.~\ref{fig:Lickpds144_2}) a trail of Herbig-Haro objects (HH 689a, b, c, d, e) is revealed. The trail extends to the northwest and southeast of PDS 144 at a position angle of -57$\degree$ and 123$\degree$ respectively (measured east of north). This jet crosses our entire field of view (13$\arcmin$ visible extent diagonally), and its total extent may well be much larger.   No significant excess emission was detected in the [SII] filter.


\subsection{Goddard Fabry-P\'{e}rot}

\begin{figure}[H]
\includegraphics[scale=1]{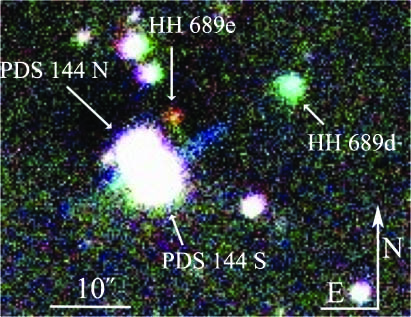}
\caption{\footnotesize This H$\alpha$ velocity scan image from the Goddard Fabry-P\'{e}rot instrument on the ARC 3.5m at Apache Point Observatory illustrates the radial velocity of the bi-polar jets of the PDS 144 stars, and HH 689d \& 689e. The 3-color image is composed of \it Blue: \rm H$\alpha$ - 114 $\pm$ 60 km s$^{-1}$ \it Green: \rm H$\alpha$ $\pm$ 60 km s$^{-1}$ \it Red: \rm H$\alpha$ + 114 $\pm$ 60 km s$^{-1}$. HH 689e lacks any blue component and is therefore pointed away from us while HH 689d lacks any red component and is therefore pointed toward us. This image covers approximately the FOV of the center inset in Fig.~\ref{fig:Lickpds144_2}.  The full FOV of the GFP is 3.7$\arcmin$, but only the central region seen here contained radial velocity data on the jets and HH knots associated with PDS 144 N \& S. \normalsize \label{fig:APO3colorPDS144v2}}	
		
\end{figure}

PDS 144 and HH 689d \& 689e were observed on the Astrophysical Research Consortium (ARC) 3.5m telescope at Apache Point Observatory using the Goddard Fabry-P\'{e}rot (GFP).  A velocity scan imaging sequence was carried out on 2008 July 6 (MJD 54653) across H$\alpha$ (6563\AA) using a high resolution etalon with a FWHM of 120 km s$^{-1}$ in five velocity steps across the outflow covering a FOV of 3.7$\arcmin$.  The observations were made under conditions of 1.5\arcsec\ seeing.  The five velocity steps span from -228 km s$^{-1}$ to +228 km s$^{-1}$ in 114 $\pm$ 60 km s$^{-1}$ steps across H$\alpha$.  GFP data reduction is discussed in \citet{Wassel_06}.  Only the stars were detected in the $\pm$228 km s$^{-1}$ frames, so the color composite image  (Fig.~\ref{fig:APO3colorPDS144v2}) contains data from the -114 km s$^{-1}$ (blue component), 0 km s$^{-1}$ (green component), and +114 km s$^{-1}$ (red component) images.


\subsection{HST Observations}

PDS 144 has been observed in the optical on four occasions with two HST instruments: the Advanced Camera for Surveys (ACS) and the Space Telescope Imaging Spectrograph (STIS).   Table~\ref{fig:HST_Obs_PDS144} provides details. 


\subsubsection{ACS}

Observations with ACS were obtained over two epochs separated by a 3.89 year baseline.  The first epoch HST optical imagery of PDS 144 N and PDS 144 S was obtained under program HST-GO-10603 on 2006 April 11 (MJD 53836) with the ACS Wide Field Channel (WFC) using the F555W \& F814W filters (5346/1193 \AA\ \& 8333/2511 \AA\ respectively) in 60 s \& 680 s exposures.  The second epoch ACS/WFC optical imagery of PDS 144 N \& S was obtained under program HST-GO-12016 on 2010 March 3 (MJD 55258) using the F606W filter (5907/2342 \AA ) in 5 s, 15 s, \& 280 s exposures.  

\begin{figure}[H]
\includegraphics[scale=0.7]{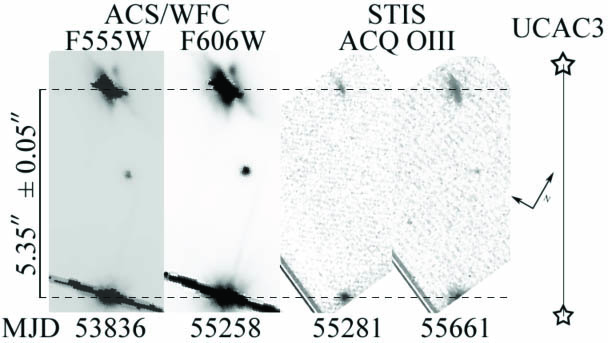}
\caption{\footnotesize The position of PDS 144 N (\it top\rm ) with respect to PDS 144 S (\it bottom\rm ) over a 5 year baseline is seen in these HST images, and they show no change in separation.  The cartoon labeled UCAC3 shows the predicted location the stars in our fourth epoch (MJD 55661) if we were to apply the UCAC3 proper motions to the two PDS 144 stars in our first epoch (MJD 53836).\normalsize \label{fig:RELPMFIGURE1}}
\end{figure}

\begin{figure}[H]
\includegraphics[scale=0.6]{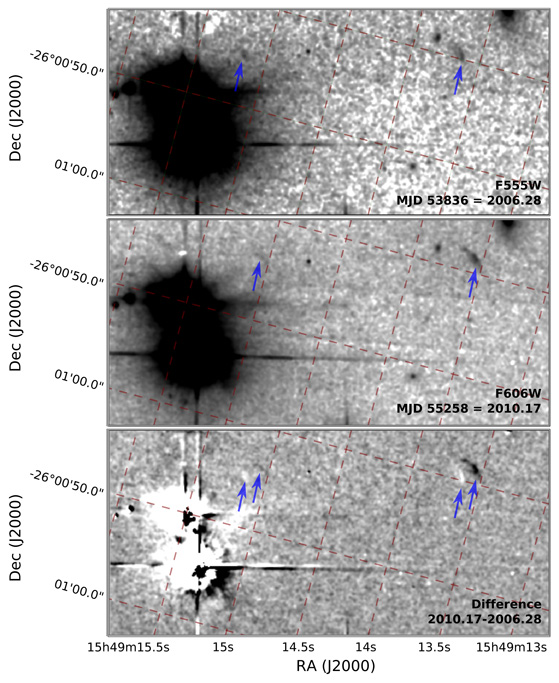}
\caption{\footnotesize This figure illustrates the relative proper motion of HH 689d \& 689e. \it Top\rm : The F555W image from ACS/WFC with PDS 144 N \& S on the left of each frame, and the location of HH 689d \& 689e is indicated in each frame by $\longrightarrow$\ , \it Middle\rm : The F606W image from ACS/WFC, and \it Bottom\rm : The F606W image with the F555W image subtracted to show the relative proper motion of HH 689e and HH 689d over the 3.89 year baseline. \normalsize \label{fig:PDS144_KnotsPM}}
\end{figure}

ACS has two separate CCD chips, both of which are 4096 $\times$ 2048 pixels with a 50 pixel gap between them. Both chips cover a total FOV of 3.3$\arcmin$.  PDS 144 N \& S were imaged in the WFC1 chip near the (sci,2) position, a point in the chip with minimal geometric distortion.  Imaging with HST's Wide Field Camera 3 (WFPC3) was an available option at the time of our second epoch observations.  The reason for using the same camera and aperture is to ensure that differences in geometric distortion are minimized, and that any incomplete compensation is a systematic error applicable to both epochs (Fig.~\ref{fig:RELPMFIGURE1} \& \ref{fig:PDS144_KnotsPM}). The initial data reduction for both epochs of data was performed by the ACS pipeline \citep{ACS_data_handbook, ACS_instrument_handbook}.  The F606W data were further processed to remove striping that appears in some data after service mission 4 \citep{STScI_ISR_ACS_2011_05}.


\subsubsection{STIS}

Target Acquisition imagery from STIS was taken over two epochs with a $\approx$ 1 year baseline.  Acquisition imagery for UV spectra of PDS 144S was obtained using the STIS CCD camera and the F28X50OIII blocking filter (5005.8/6.2 \AA). The field of the acquisition imagery is 5$\arcsec$ x 5$\arcsec$, with a pixel scale of 0.05$\arcsec$ pixel$^{-1}$.  This is sufficient to image both PDS 144 stars simultaneously at the nominal separation noted by \citet{Perrin_06} with the +y axis aligned along PA = -114.588$\degree$ (east of north). The acquisition data comprise short exposures to locate the star prior to positioning in the small STIS spectrographic slits. With data obtained on 2010 March 26 (MJD 55281) and 2011 April 10 (MJD 55661), also as part of HST-GO-12016, the STIS ACQ observations provide two additional epochs of optical imagery at HST's angular resolution suitable for measuring the proper motion of the PDS 144 stars (Fig.~\ref{fig:RELPMFIGURE1}). The data were processed using the STIS pipeline software (CALSTIS version 2.32 or later in STSDAS version 3.13) and on-the-fly calibrations as discussed in \citet{STIS_instrument_handbook} \& \citet{STIS_data_handbook}.  The spectroscopic measurements obtained with STIS following these acquisition images will be discussed in a future paper.

\section{ANALYSIS \& RESULTS}

\subsection{Jets}

\subsubsection{Jet Orientation}

PFCam data first indicated the presence of bipolar jets from both PDS 144 N \& S by revealing HH 689 extending 5$\arcmin$ along PA = 303$\degree$ $\pm$ 7 $\degree$ and 8$\arcmin$ along PA = 123$\degree$ $\pm$ 7 $\degree$ (distance measured from PDS 144, and all angles measured east of north). HH 689e \& HH 689d are the closest HH knots to PDS 144 N \& S (8\arcsec\ and 20\arcsec\ respectively) and seen in the center inset of the PFCam image (Fig.~\ref{fig:Lickpds144_2}), and are close enough to be observed in both the GFP \& HST/ACS imagery of PDS 144.  The GFP data (Fig.~\ref{fig:APO3colorPDS144v2}) confirm the presence of the bipolar jets, and the high resolution of the HST imagery further constrains the alignment of the two jets. The angle from PDS 144 N to HH 689e  is 311$\degree$ $\pm$ 4$\degree$ and the angle from PDS 144 S to HH 689d is 304$\degree$ $\pm$ 2$\degree$, therefore they are aligned to within 7$\degree$ $\pm$ 4.7$\degree$ on the sky.  The PA of HH 689 from PDS 144 N is seen to be orthogonal to that of the disk midplane, which lies along PA = 45$\degree$ $\pm$ 5$\degree$.

\subsubsection{Jet Radial Velocity \& Proper Motion}

The GFP velocity scan color composite image across H$\alpha$ (Fig.~\ref{fig:APO3colorPDS144v2}) provides us with radial velocity information on the jets/HH knots associated with PDS 144.  HH 689e was detected in both the +114 km s$^{-1}$ and 0 km s$^{-1}$ frames, and HH 689d was detected in both -114 km s$^{-1}$ and 0 km s$^{-1}$ frames.  This demonstrates that, while the jets are closely aligned on the sky, they are oppositely inclined towards us.  These observations were not made under ideal conditions (1.5\arcsec\ seeing), so we have used the velocity midway between 0 km s$^{-1}$ and $\pm$ 114 km s$^{-1}$ in estimating the radial velocity of the HH knots.  HH 689e is moving away from us at +57 $\pm$ 60 km s$^{-1}$, and HH 689d is moving toward us at -57 $\pm$ 60 km s$^{-1}$.  

We have measured the proper motion for HH 689d \& 689e using the HST/ACS optical observations spanning a 3.89 year baseline. The proper motion (measured with respect to the star) for HH 689e is 480 $\pm$ 70 mas/yr, and the proper motion for HH 689d is 400 $\pm$ 30 mas/yr (Fig.~\ref{fig:PDS144_KnotsPM}).  HH 689e is more difficult to resolve in the imagery because it is seen amidst the reflection nebulosity of the outflow cavity from PDS 144 N, and thus its proper motion carries larger uncertainty. Assuming no further deceleration of the outflows, these proper motion data imply that HH 689a is at least 990-1210 years old.  The maintenance of aligned outflows over that (or longer) time intervals suggests that the PDS 144 Herbig Ae stars are comoving.


\subsection{Relative Motion of Stars}

We measure the separation between PDS 144 N \& S to be 5.35$\arcsec$ $\pm$ 0.05$\arcsec$ in each of the four epochs spanning five years.  However, the proper motion reported in the UCAC3 \citep{Zacharias_10} indicates PDS 144 N is moving in the opposite direction as PDS 144 S (see Table~\ref{fig:PMtablePDS144} \it bottom\rm ).  Were this the case, given their reported measurements, over a 5 yr baseline between the first and fourth epochs we should have seen an increased separation between PDS 144 N \& S of 0.80$\arcsec$ $\pm$ 0.21$\arcsec$.  Analysis of our multi-epoch imagery taken over a 5 year baseline (Fig.~\ref{fig:RELPMFIGURE1}) demonstrates that no change in separation occurred between observations, so we conclude that the UCAC3 data for PDS 144 is in error at the 7$\sigma$ level.

\subsection{PDS 144 are Members of Upper Sco}

Precise measurements of the proper motions of the PDS 144 stars allow us to test their dynamical membership in nearby young stellar associations.  Conveniently, there is a prominent spiral galaxy visible 28\arcsec\ northeast of PDS 144, providing an absolute reference for zero proper motion.   We measured proper motions for the PDS 144 stars and these field sources by comparing the position of stars as well as the spiral galaxy in the F555W \& F814W images with that seen in the F606W image.  Differences in position between epochs due to errors in pointing were then eliminated by setting the proper motion of the spiral galaxy to 0.  Comparison of the two epochs shows that the only sources with significant proper motion are PDS 144 N and S and the nearby T-Tauri star TYC 6782-878-1 (Table 2 and Fig. 5).   For PDS 144 N we find proper motion ($\mu _{\alpha} \cos \delta, \mu _{\delta}$) =  (-20, -22) $\pm$ 10.5 mas yr$^{-1}$, and for PDS 144 S we find (-17,-22) $\pm$ 10 mas yr$^{-1}$.   This is consistent with the mean proper motion of Upper Sco, ($\mu_{\alpha} \cos \delta , \mu_{\delta}$) = (-25, -10) mas yr$^{-1}$, as reported by \citet{DeZeeuw_99}. We conclude that PDS 144 N \& S are members of the Upper Sco association. We therefore adopt the  mean distance estimate for Upper Sco of 145 $\pm$ 2, where the error is the formal error on the mean \citep{DeZeeuw_99}.  The Upper Sco region spans $\approx$ 14$\degree$ on the sky which at 145 pc distant corresponds to a transverse size of $\approx$ 36 pc.  If we assume the region extends along our line of sight a similar distance, then the distance to PDS 144 is 145 $\pm$ 18 pc.



Roughly 1.5\arcmin\ away from PDS 144 lies the weak-lined T-Tauri system TYC 6782-878-1 (also referred to as [PZ99] J154920.9-260005 and RX J1549.3-2600), which is a known member of the Upper Sco association \citep{Preibisch_99}, subgroup Upper Sco A \citep{Kohler_00}.   This system has been resolved into a 0.16\arcsec\ binary with a flux ratio 0.5 at K band \citep{Kohler_00, Woitas_01}.  Like some 90\%\ of weak T-Tauri stars, its spectral energy distribution shows no evidence of infrared excess from dust (L$_{disk}$/L$_{star} <$ 8.8 $\times$ 10$^{-8}$; \citet{Padgett_06, Wahhaj_10}).  Its proper motion is identical within the uncertainties with that of PDS 144, and at a distance of 145 pc, its projected separation from PDS 144 is 13000 AU = 0.06 pc.

Based on a comparison of that star's luminosity to the stellar evolutionary tracks of \citet{Siess_00}, \citet{Wahhaj_10} estimated an age of 11 Myr for TYC 6782-878-1.  The age of the Upper Sco region is typically taken to be 5 $\pm$ 1 Myr \citep{Preibisch_99}, but recent work suggests it may indeed be closer to 10 Myr (Pecaut \& Mamajek, in prep). We thus adopt a range of 5 - 10 Myr as our best estimate for age of the PDS 144 stars.

All other sources in the field of view show proper motions consistent with being background objects. In particular, almost precisely between the two PDS 144 stars lies a much fainter point source, which Perrin et al. (2006) suggested was most likely an unassociated background star. Our proper motion measurements now confirm this hypothesis, showing it to have a near-zero proper motion that is inconsistent with the PDS 144 stars at a 5$\sigma$ level. SED fits to the V to K band photometry of this star indicate it can be well fitted by a mid-K main sequence star approximately 1700 pc distant. 

\section{DISCUSSION}

PDS 144, with an A2Ve primary and an A5Ve secondary, is the first confirmed example of a nearly equal-mass Herbig Ae binary with disks which are sufficiently spatially separated that they can be studied with the full range of techniques devoted to disks around single stars.  They are also physically close enough to study any history of dynamical interaction between the two star+disk systems.  This makes the stars of PDS 144 unique among the known Herbig Ae stars, and they will serve as a laboratory for comparative disk structure and chemistry studies. 

\citet{Perrin_06} derived properties for the PDS 144 binary and the disk of PDS 144 N assuming a distance of 1 kpc. They measured the dark lane to be $\approx$0.8$\arcsec$ across and 0.15$\arcsec$ in height.  Using our revised distance of 145 $\pm$ 2 pc, these correspond to a disk radius of 58 $\pm$ 13 AU and a disk height of 22 $\pm$ 2 AU for PDS 144 N.  This disk radius is smaller than that of many of the Herbig Ae disks, for instance the AB Aur or HD 100546 disks which are many hundreds of AU across.  It may be comparable in size to that of HD 104237, for which \citet{Grady_04} placed an upper limit on the disk radius of r $\leq$ 70 AU. The most likely explanation is dynamical truncation from interactions between the two PDS 144 stars.  Our measured separation of 5.35$\arcsec$ $\pm$ 0.05$\arcsec$ \citep{Perrin_06} between PDS 144 N \& S corresponds to a physical separation of $\geq$ 771 $\pm$ 7.5 AU.  It is interesting to note that with a radius of 58 AU, the disk around PDS 144 N is only slightly larger in radius than the radius of our own Solar System's Kuiper belt, $\approx$50 AU.  

\subsection{Disk Inclination of PDS 144 S}

\citet{Perrin_06} suggested PDS 144 S might be viewed at low inclination due to the absence of any disk detection in HST and Keck imagery.  If we assume, however, that both stars are driving jets which are orthogonal to the disk (true for PDS 144 N), we  can use GFP radial velocity measurements and the proper motion data of HH 689d to estimate its jet inclination and thus place constraints on its disk inclination.  For PDS 144 S, combining results from \S\ 3.1.2 for the radial velocity from the GFP of -57 $\pm$ 60 km s$^{-1}$ and the velocity equivalent at 145 pc of the proper motion of 275 $\pm$ 20 km s$^{-1}$, we calculate a disk inclination of i = 78$\degree$ $\pm$ 12$\degree$. However, the dark lane of PDS 144 N is clearly observed at all wavelengths and has an inclination of $\approx$83$\degree$ \citep{Perrin_06}.  This implies, due to the complete absence of a dark lane in the direct imagery of PDS 144 S, that its disk is seen at a lower inclination than PDS 144 N. We have therefore placed an upper limit of 80$\degree$ inclination on the disk of PDS 144 S, narrowing the inclination range to i = 73$\degree$ $\pm$ 7$\degree$.  Similarly, using the radial velocity of HH 689e (+57 $\pm$ 60 km s$^{-1}$; see \S\ 3.1.2) with the proper motion velocity equivalent (330 $\pm$ 50 km s$^{-1}$; see \S\ 3.1.2), we arrive at an inclination range of i = 79$\degree$ $\pm$ 11$\degree$ for the disk around PDS 144 N. This agrees quite well with the i = 83$\degree$ $\pm$ 1$\degree$ derived for the disk of PDS 144 N by \citet{Perrin_06}.

Also easily calculated with a distance of 145 pc, and the results above, are the space velocities for HH 689d \& 689e which are 281 $\pm$ 63 km s$^{-1}$ and 335$\pm$ 78 km s$^{-1}$ respectively.  While the literature suggests that the typical jet outflow velocity for Herbig Ae stars is in a range between 200 km s$^{-1}$ and 400 km s$^{-1}$ \citep{Mundt_87, McGroarty_04, McGroarty_07}, there is only limited observational basis to date for such an assumption.  The space velocities of HH 689d \& 689e provide two additional data points to substantiate this velocity range. 

\subsection{Disk Alignment and Comparison with T-Tauri Wide Binaries}

The jets of PDS 144 N \& S are closely aligned on the sky, but the approaching jet components for the stars are oppositely directed, so the inclination of the disks relative to one another is 25$\degree$ $\pm$ 9$\degree$. \citet{Jensen_04} demonstrated that disk orientations in T-Tauri wide binaries older than 1 Myr should have inclinations that appear aligned to within 20$\degree$ $\pm$ 10$\degree$ relative to one another.  However, the work by \citet{Jensen_04} was based solely on polarimetric observation, and Jensen noted that polarization studies by themselves also lack the ability to determine which side of the disk is inclined toward or away from us.  Therefore our measurement of the relative disk alignment for PDS 144 is in agreement with Jensen's study.  This also suggests we are seeing the binary plane of PDS 144 nearly edge-on.  

In a more recent study, \citet{Wheelwright_11} take into account the possible projection effects that limited the certainty of the results in \citet{Jensen_04}.  The spectropolarimetric study of Herbig Ae/Be binaries by \citet{Wheelwright_11} confirms the results in \citet{Jensen_04} by concluding that random association of the circumprimary disk and binary plane can be excluded.  Wheelwright's observations consisted of Herbig Ae/Be binaries with T-Tauri companions, and yielded results similar  to those in \citet{Jensen_04}. This study provides further confirming evidence that disks around Herbig Ae/Be and T-Tauri stars in binary systems are coplanar with the binary plane of the system.  We measure the PA from PDS 144 S to PDS 144 N to be 29$\degree$ $\pm$ 1$\degree$ and the PA of the disk midplane of PDS 144 N to be 45$\degree$ $\pm$ 5$\degree$.  If we assume we are seeing the binary plane of PDS 144 N \& S edge-on, the result that PDS 144 N is within 16$\degree$ $\pm$ 5$\degree$ of being aligned with the binary plane is consistent with and could further confirm Wheelwright's hypothesis.

\subsection{Disk Geometry}

The majority of known Herbig star binaries have lower mass, weak  T-Tauri companions \citep{Leinert_97, Baines_06}. Those less massive companions, however, have typically shown no evidence of retaining any circumstellar disk material \citep{Hall_97, Grady_05}.  PDS 144 S, being nearly equal in mass to PDS 144 N, seems to have managed to hold on to its disk but the disk shows distinct spectral differences to that of PDS 144 N. \citet{Perrin_06} detected polycyclic aromatic hydrocarbon (PAH) emission in the disk of N, while the disk of S lacks PAH emission. 

There are two requirements for PAH emission: (1) PAHs must be present in the disk, and (2) they must be UV irradiated.  The disk around PDS 144 N is flared, which is favorable for illumination of the optically thick dust disk surface. So the simplest explanation for why the disk around PDS 144 S lacks PAH emission is that interaction with PDS 144 N has sufficiently flattened and stratified the disk such that any PAHs present in the disk escape UV irradiation.  Geometrical differences in the disk could be entirely responsible for the spectral differences seen between N \& S. PSF template star subtraction from PDS 144 S is one method that may allow the detection of its disk directly and make study of its disk's geometry accessible (Hornbeck et al. 2011, in prep).

Circumstellar disks at inclinations in the range inferred from the jet measurements are not typically visible in direct light imagery since the star, while reddened, may not be occulted by the disk (as is the case of PDS 144 N).  This inclination range for PDS 144 S is where large-amplitude light and polarization variability (UXOR behavior) is expected \citep{Natta_97}.  Photometric observations of PDS 144 S will reveal whether this is in fact the case (Hornbeck et al. 2011, in prep).

\section{CONCLUSIONS}

PDS 144 N \& S are comoving and have a common proper motion with TYC 6782-878-1, a T-Tauri star with association membership in Upper Sco.  Association membership of PDS 144 with Upper Sco establishes an age of t = 5 - 10 Myr and distance of d = 145 $\pm$ 2 pc, and makes PDS 144 N \& S the first confirmed Herbig Ae - Herbig Ae binary with disks which can be spatially resolved and studied as if they were disks around single stars.  The inclinations of the disks around PDS 144 N \& S, relative to one another, are similar to those seen around T-Tauri wide binaries.  The two PDS 144 stars are also both being seen at high inclination, and they are among a small set of Herbig Ae stars known to drive jets associated with parsec-scale chains of HH knots.  Their age, however, is not typical of other Herbig Ae stars that have been observed driving jets.

\section{ACKNOWLEDGEMENTS}

This work, in part, is based on observations made with the NASA/ESA Hubble Space Telescope, obtained at the Space Telescope Science Institute, which is operated by the Association of Universities for Research in Astronomy, Inc., under NASA contract NAS 5-26555. These observations are associated with programs HST-GO-11155, HST-GO-12016, Chandra program 11200435 and NASA RTOP 399131.02.02.02.32 to Goddard Space Flight Center. Data used in this study were also  obtained under HST-GO-10603.  Data used in this study were also obtained at the Lick Observatory Shane 3 m telescope, the Apache Point Observatory 3.5 m, and the Keck Observatory. The Apache Point observations were made under a grant of Director's Discretionary Time.  J.P.W. is supported by NSF Astronomy \& Astrophysics Postdoctoral Fellowship, AST 08-02230.  M.D.P. was supported in part by an NSF Astronomy \& Astrophysics Postdoctoral Fellowship, AST-0702933. This research was supported in part by a grant from the Kentucky Science and Engineering Foundation as per Grant Agreement \# KSEF-148-502-08-241 with the Kentucky Science and Technology Corporation and NASA Kentucky Space Grant Consortium Awd \# 3049024102-11-175. D.G.B. was partly supported by a NASA Postdoctoral Program Fellowship administered by Oak Ridge Associated Universities.

\bibliographystyle{apj}

\begin{figure}[H]
\includegraphics[scale=0.45]{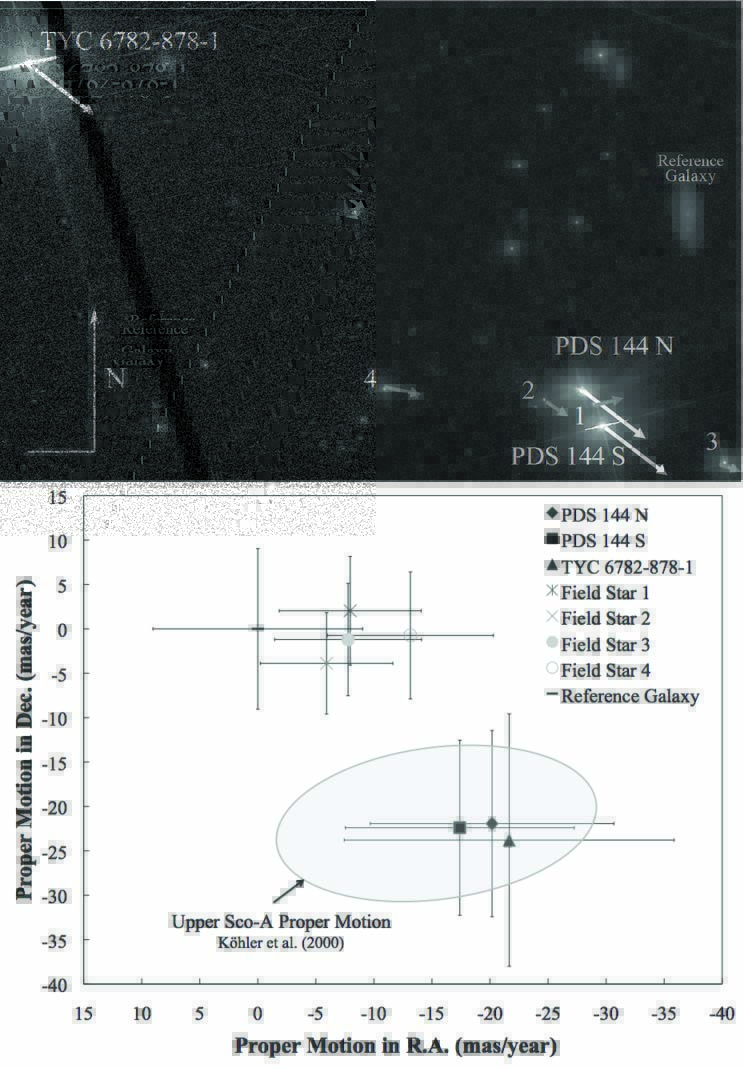}  
\caption{\footnotesize \it Top\rm : ACS/WFC F555W image of PDS 144. Arrows (not to scale) show the magnitude and direction of proper motion for PDS 144 N, PDS 144 S, \& TYC 6782-878-1 as well as that of background objects seen nearby in the field: 1 - uncatalogued field star (\it seen directly between PDS 144 N \& S\rm ), 2 - uncatalogued field star, 3 - 2MASS 15491411-2600598, 4 - 2MASS 15491748-2600499. \it Bottom\rm : Chart illustrating the proper motions of the stars seen in the ACS/WFC F555W image above.  The grey ellipse encompasses the proper motion of stars in Upper Sco A. It was generated from the list of Upper Sco A stars in \citet{Kohler_00} that also had proper motion measurements available in the literature.  Note: \it Error bars are 3$\sigma$. \normalsize \label{fig:ProperMotionPDS144_v4}}
\end{figure}

\begin{deluxetable}{llllll}
\tablecolumns{5}
\tablewidth{0pt}
\tablecaption{HST Optical Observations of PDS 144 \label{fig:HST_Obs_PDS144}}
\tablehead{

\colhead{Dataset\tablenotemark{a}}    &  \colhead{Date}   & \colhead{Instrument}  & \colhead{Filter} & \colhead{Exp (s)}}
\startdata
J9F509011         &   2006 April 11 & ACS/WFC   & F555W      &    740  \\
J9F509021         &   2006 April 11 & ACS/WFC   & F814W      &    740  \\
JBDF08011         &   2010 March 3  & ACS/WFC   & F606W      &    300  \\
OBDF09WFQ         &   2010 March 26 & STIS/ACQ  & F28X50OIII &    \phantom{0}52.1  \\ 
OBDF04R3Q         &   2011 April 10 & STIS/ACQ  & F28X50OIII &    200.1  \\ 
\enddata
\tablenotetext{a}{\ Reference number for dataset within HST archive}
\end{deluxetable}

\begin{deluxetable}{lr@{\ $\pm$\ }lr@{\ $\pm$\ }l}
\tablecolumns{5}
\tablewidth{0pc}
\tablecaption{Stellar Proper Motion Results\tablenotemark{a}\label{fig:PMtablePDS144}}
\tablehead{
\multicolumn{1}{l}{Object} &\multicolumn{2}{c}{$\Delta$RA (mas yr$^{-1}$)}    &\multicolumn{2}{c}{$\Delta$Dec. (mas yr$^{-1}$)} }
\startdata
PDS 144 N             & -20 & 10.5                 & -22 & 10.5                   \\
PDS 144 S             & -17 & 10                   & -22 & 10                     \\
TYC 6782-878-1        & -22 & 14                   & -24 & 14                     \\
S9DZ000424            & -8  & 6                    &  2  & 6                      \\
Field Star 2          & -6  & 6                    & -4  & 6                      \\
2MASS15491411-2600598 & -8  & 6                    & -1  & 6                      \\
2MASS15491748-2600499 & -13 & 7                    & -1  & 7                      \\
\cutinhead{UCAC3 Suggested Proper Motions\tablenotemark{b}}
PDS 144 N             & 24.0 & 27                    & 41.6 & 27                    \\
PDS 144 S             & -57.3 & 20                 & -96.6 & 20                   \\
TYC 6782-878-1        & -17.7 & 34                 & -29.1 & 34                   \\
\enddata
\tablenotetext{a}{\footnotesize Proper motion measurements for stars seen in the HST/ACS imagery of PDS 144.}
\tablenotetext{b}{\footnotesize Proper motions given in UCAC3. While our data for PDS 144 N \& S disagree, values from both datasets for TYC 6782-878-1 are in agreement with one another.}
\end{deluxetable} 

\end{document}